\documentclass{svmult}

\usepackage{makeidx}         
\usepackage{graphicx}        
\usepackage{multicol}        
\usepackage[bottom]{footmisc}

\makeindex             

\begin{document}
 
\title*{Fractal Stationary Density in Coupled Maps}
\author{J\"urgen Jost \and
Kiran M. Kolwankar}
\authorrunning{Jost and Kolwankar} 
\institute{Max Planck Institute for Mathematics in the Sciences,
Inselstrasse 22-26, 
D-04103 Leipzig, Germany}

\maketitle

\begin{abstract}
We study the invariant measure or the stationary density of a
coupled discrete dynamical system as a function of the coupling
parameter $\epsilon$ ($0 < \epsilon < 1/4$). The dynamical system
considered is chaotic and unsynchronized for this range of parameter
values. We find that the stationary density, 
restricted on the synchronization manifold, is a fractal function.
We find the lower bound on  the fractal dimension of the graph 
of this function and show that it
changes continuously with the coupling
parameter.
\end{abstract}

\section{Introduction}
\label{sec:1}
Two or more coupled rhythms can under certain conditions synchronize,
that is a definite relationship can develop between these rhythms.
This phenomena of synchronization~\cite{PRK,Pecora1997}
in coupled dynamical systems has acquired immense importance in
recent years since it appears in natural phenomena as well as in
engineering applications. 
What is even more interesting is the fact
that even chaotic oscillations
can synchronize. This observation is being
utilized in secure communication~\cite{VanWiggeren1998}.
Studying synchronization is also important for neural information
processing ~\cite{Hansel1992,Buzsaki2004,Friedrich2004}.

These developments have lead to the investigation of coupled dynamical systems.
The dynamical systems considered can either be continuous or
discrete in time. Different types of couplings have been considered,
like continuously coupled or pulsed coupled. And many 
coupling topologies arise in nature as well as in human applications, like random,
all-to-all, scale-free, small-world, nearest neighbour lattice, etc.
Also the coupling strengths can vary from element to element.
We plan to study a system of two coupled maps which forms a 
basic building block of all these systems and allows us to
separate the complexity due to coupling topologies from
that coming from the chaotic nature of the dynamical system
considered. As a next step, one can then consider various
coupling topologies. We have demonstrated in~\cite{Jost2004}
that the result of such a system of two coupled maps can be
used to derive the result for a globally coupled network of $N$ maps. 
Here we are concerned with the phenomena of complete synchronization,
that is, the dynamics of two systems becomes completely identical
after the coupling parameter crosses a certain critical value.

\paragraph{The model}
We consider the
following coupled map system
\begin{eqnarray}\label{eq:system}
X_{n+1}=AF(X_n) := S(X_n)
\end{eqnarray}
where $X=(x,y)^T$ is a 2-dim column vector, $A$ is a $2\times 2$ 
coupling matrix and $F$ is a map from $\Omega=[0,1]\times [0,1]$ onto
itself. In the present paper we take $F$ to be the extension of
the tent map $f_t:[0,1]\rightarrow [0,1]$,
\begin{eqnarray}\label{eq:tent}
f_t(x) = \left\{ \begin{array}{lcl}
2x && 0\leq x \leq 1/2 \\
2-2x && 1/2 \leq x \leq 1
\end{array} \right. ,
\end{eqnarray}
to two variables and we choose
\begin{eqnarray}
A=\left( \begin{array}{cc}
1-\epsilon & \epsilon \\
\epsilon & 1-\epsilon
\end{array} \right)
\end{eqnarray}
where $0< \epsilon<1$ is the coupling strength.
This type of coupling has been called contractive or dissipative.
See \cite{PRK} for the physical motivation behind
considering such a system.
Furthermore, the row sums of $A$ are equal to one which
guarantees the existence of a synchronized solution.

\begin{figure}
\centering
\includegraphics[height=4cm]{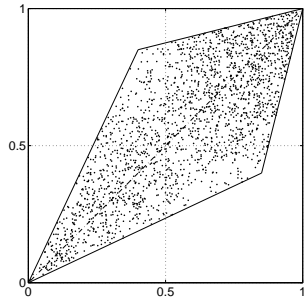}
\caption{Asymptotic distribution of 2000 points starting from
an uniform distribution over the whole phase space $\Omega$.
$\epsilon=0.2$.}
\label{fig:1}       
\end{figure}
Before proceeding we give a mathematical definition of synchronization
we are interested in.
\begin{definition}
A discrete dynamical system $S:\Omega\rightarrow\Omega$ is said 
to completely synchronize if as $n$ tends to infinity
$|S^n(x)-S^n(y)|$ tends to zero for all $(x,y)\in\Omega$.
\end{definition}

\begin{figure}
\centering
\includegraphics[height=4cm]{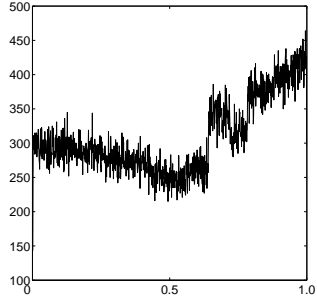}
\caption{The section of the stationary density along the line $x=y$.
In this graph the total number of points used is $10^8$ and the size
of the mesh used is 1000x1000.}
\label{fig:3}      
\end{figure}

Using the linear stability analysis it can be shown~\cite{jost2002}
 that this coupled map system
synchronizes when $1/4 < \epsilon < 3/4$. The same result was also
obtained by studying the evolution of the support of the invariant
measure~\cite{Jost2004} which is a global result as opposed to the
linear stability analysis which is carried out near the synchronized
solution. This was done by showing that if $\epsilon < 1/4$
we obtain a quadrilateral of nonzero area as the support of the
invariant measure. We depict this area in Fig~\ref{fig:1} along with a
distribution of points obtained from uniform distribution of
initial conditions. And when $\epsilon$ crosses the value 1/4
this quadrilateral shrinks to the line $x=y$. In this sense the
synchronization transition is a discontinuous transition.
\begin{figure}
\centering
\includegraphics[height=4cm]{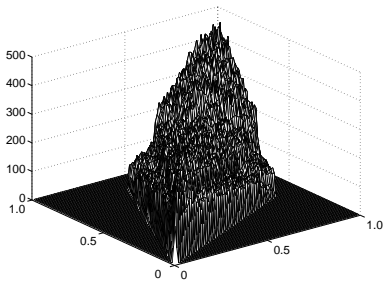}
\caption{The stationary density: a histogram of the distribution of
points of Fig.~\ref{fig:1}. Here the number of point is $10^6$
and the mesh size is 100x100. The horizontal plane depicts $\Omega$
and the vertical axis gives the number of points lying in the mesh
square.}
\label{fig:2}      
\end{figure}
But this figure is misleading as it does not tell us anything about
the density of points. As we show in Fig.~\ref{fig:2}, if we plot the histogram
then we see a quite irregular structure. 
In Fig.~\ref{fig:3} we plot the section of this invariant density 
along the line $x=y$. We see clearly that it is an 
irregular function.
Studying this invariant
density is the object of this paper. In particular we show that
this density is indeed a fractal function on the synchronization manifold,
i.e., its section along line $x=y$, and the fractal dimension of the graph
of this function depends on the coupling parameter.

The paper is organized as follows. In section~\ref{sec:2} we give
a brief introduction to invariant measures recalling some definitions
and results required for completeness. We also outline a method
to find the stationary density. We then move on to our main result
in the section~\ref{sec:3} after introducing basic concepts needed
to characterize the irregularity of fractal functions. Section~\ref{sec:4}
concludes by pointing out some future directions.

\section{Invariant measure}
\label{sec:2}

Invariant measures or the stationary densities~\cite{LM} provide a useful
way to study the asymptotic behavior of dynamical systems. One starts with a 
distribution of initial conditions and studies its evolution as time
goes to infinity. 
It is interesting to note that even if the dynamical system
is chaotic a well behaved limit can exist which can then be used
to study various average properties. In this section we give
a brief introduction to invariant measures and a way to find
one. We begin with the definitions (the norms used are $L^1$ norms
throughout):

\begin{definition}
A measure $\mu$ is said to be invariant under a transformation
$S$ if $\mu(S^{-1}(A)) = \mu(A)$ for any measurable subset $A$ of $\Omega$.
\end{definition}

\begin{definition}
Let $(X,A,\mu)$ be a measure space and the set $D(X,A,\mu)$
be defined by $D(X,A,\mu)=\{ f \in L^1(X,A,\mu): f\geq 0
\mbox{ and } ||f||=1 \}$. Any function $f \in D(X,A,\mu)$ is called
a density.
\end{definition}

\begin{definition}
Let $(X,A,\mu)$ be a measure space. A linear operator
$P:L^1 \rightarrow L^1$ satisfying

\noindent
(a) $Pf \geq 0$ for $f\geq 0$, $f \in L^1$; and

\noindent
(b) $||Pf|| = ||f||$, for $f\geq 0$, $f \in L^1$

\noindent
is called a Markov operator.
\end{definition}
A Markov operator satisfies a {\em contractive property}, viz., 
$||Pf|| \leq ||f||$. And this property implies the {\em stability
property} of iterates of Markov operators, viz.,
$||Pf-Pg|| \leq ||f-g||$.
We shall be interested in the fixed points of  Markov operators.

\begin{definition}
Let $(X,A,\mu)$ be a measure space and $P$ be a Markov operator.
Any $f\in D$ that satisfies $Pf=f$ is called a stationary density
of $P$.
\end{definition}
A stationary density is the Radon-Nikodym derivative of an 
invariant measure with respect to $\mu$.

\subsection{The Frobenius-Perron operator}
There exist various methods to find  invariant measures. One approach
is to use the so called Frobenius-Perron operator.
This operator when applied to $\rho_n(x)$, the density at the
$n$th time step, yields the density at the $(n+1)$th time step. Since
all the points at the $(n+1)$th step in some set $D$ come from
the points in the set $S^{-1}(D)$ we have the following equality 
 defining the Frobenius-Perron operator $P$:
\begin{eqnarray}\label{eq:FP}
\int\int_D P\rho(x) \mu(dx) = \int\int_{S^{-1}(D)} 
\rho(x) \mu(dx)
\end{eqnarray}
The Frobenius-Perron operator is a Markov operator.

\subsection{The invariant measure of the tent map }
If in~(\ref{eq:FP}) we choose our discrete dynamical system to be the one
dimensional tent map defined in~(\ref{eq:tent})  and $D=[0,x]$ then
the equation~(\ref{eq:FP}) reduces to
\begin{eqnarray}
P\rho(x) = \rho(x/2) +\rho(1-x/2).
\end{eqnarray} 
We are interested in the fixed point solutions; this leads to
\begin{eqnarray}
\rho(x) = \rho(x/2) +\rho(1-x/2).
\end{eqnarray}
A solution of this functional equation is $\rho(x)=1$. Of course, this
is a trivial solution. 
$\delta(x)$ and $\delta(x-2/3)$ also solve this
equation but we are not interested in such singular 
solutions since they do not span the phase space.

\section{Stationary density of the coupled tent map} 
\label{sec:3}
We now turn to coupled maps. 
In this section we study the stationary density on the
synchronization manifold, i.e., the line $x=y$. We show
that the density is a fractal function with its H\"older
exponent related to the coupling parameter $\epsilon$.
 
We use the following definition of  H\"older 
continuity and its relation to the box dimension~\cite{falconer1990}:
\begin{definition}
A function $f:[0,1] \rightarrow R$ is in $C_{x_0}^\alpha$, for 
$0<\alpha <1$ and $x_0 \in [0,1]$, if for all $x \in [0,1]$
\begin{equation}\label{eq:holder}
|f(x) - f(x_0)| \leq c|x-x_0|^\alpha .
\end{equation}
\end{definition}
A {\em pointwise H\"older exponent} $\alpha_p(x_0)$ at $x_0$ is the
supremum of the $\alpha$s for which the inequality~(\ref{eq:holder}) holds. 
We also use the relation between the H\"older continuity
and the box dimension of the graph a function,
${\rm dim_Bgraph}f$.
\begin{proposition}
If for $f:[0,1] \rightarrow R$, the pointwise H\"older
exponent is $\alpha$ for all
$x\in [0,1]$ and $c$ in~(\ref{eq:holder}) is uniform
 then ${\rm dim_Bgraph}f = 2-\alpha$.
\end{proposition}
Now we are ready to state and prove our main result, viz,
the estimate of the box dimension of the graph of the 
stationary density.
\begin{theorem}
Let $\rho(x,y)$ be the stationary density of the coupled
dynamical system~(\ref{eq:system}) and let $\rho^D(x)$ be
its restriction on the line $x=y$. 
If $\rho(x,y)$ is bounded then ${\rm dim_Bgraph}\rho^D \geq d$ 
where
$d=2+\ln(1-2\epsilon)/\ln 2$, with $0<\epsilon \leq 1/4$.
\end{theorem}
{\em Proof:} We use the Frobenius-Perron operator defined in 
equation~(\ref{eq:FP}).
We choose $D=[0,x]\times[0,y]$ and get
\begin{eqnarray}
P\rho(x,y)  ={\partial\over\partial x}
{\partial\over\partial y} \int\int_{S^{-1}(D)} 
\rho(x',y') dx'dy'
\end{eqnarray}
Our $S$ in equation~(\ref{eq:system}) is not invertible. 
In fact, it has 4 preimages.
Let us denote them by $S_i^{-1}$, $i=1,...,4$. If $X\in\Omega$,
since $f$ is symmetric, we get 
\begin{eqnarray}
P\rho(X) = J^{-1}(X) \sum_{i=1}^{4} \rho(S_i^{-1}(X))
\end{eqnarray}
where $J^{-1}(X)=|dS^{-1}(X)/dX|$.
The fixed point of this operator is given by
the following functional equation for 
the density.
\begin{eqnarray}
\rho(x,y)&=&{1\over{4|1-2\epsilon |}}[
\rho(\beta x/2-\gamma y/2, -\gamma x/2+\beta y/2)
\nonumber \\
&&+\rho(1-\beta x/2+\gamma y/2, -\gamma x/2+\beta y/2)
\nonumber \\
&&+\rho(\beta x/2-\gamma y/2,1+\gamma x/2-\beta y/2)
\nonumber \\
&&+\rho(1-\beta x/2+\gamma y/2,1+\gamma x/2-\beta y/2)
]
\end{eqnarray}
where $\gamma=\epsilon/1-2\epsilon$ and $\beta = 1+\gamma$.
Since we know that a point belonging to $\Omega$ does not
leave $\Omega$, all the arguments of $\rho$ on the right hand
side of the above equation should be between 0 and 1. This gives
us four lines 
which bound an area 
$0\leq \beta x/2-\gamma y/2 \leq 1$ and 
$0\leq -\gamma x/2+\beta y/2 \leq 1$.
Lets denote this area by $Gamma$.
The support of the invariant measure should be 
contained in $\Gamma \cap \Omega$.

We also remark that for $0\leq \epsilon < 1/4$, the discrete
dynamical system $S$ that we have considered is everywhere
expanding and this implies that the stationary density 
exists~\cite{LM}.

The above equation can be written as, for $0\leq \epsilon < 1/4$,
\begin{eqnarray}\label{eq:symfnl}
\rho(x,y)={1\over{(1-2\epsilon )}}
\rho_{SS}(\beta x/2-\gamma y/2, -\gamma x/2+\beta y/2)
\end{eqnarray}
where
\begin{eqnarray}
\rho_{SS}(x,y) = {{\rho(x,y)+\rho(1-x,y)+\rho(x,1-y)+\rho(1-x,1-y)}\over 4}
\end{eqnarray}
is the part of $\rho$ that is symmetric around $1/2$ in both  arguments.
Now if we substitute $x=y$ in equation~(\ref{eq:symfnl}) we see that
the arguments on both sides of the equation belong to the diagonal. 
As a result we obtain a functional equation 
\begin{eqnarray}
\rho^D(x)={1\over{(1-2\epsilon )}}
\rho^D_{S}(x/2)
\end{eqnarray}
where we use a shorthand notation $\rho^D_{S}(x)=\rho^D_{SS}(x,x)$.
With the change of variable $z=x/2$ and a decomposition of $\rho^D(x)$ 
as $\rho^D(x) = \rho^D_S(x)+\rho^D_A(x)$, the "symmetric" and "antisymmetric"
part where again $\rho^D_A(x)$ is a shorthand notation for
$\rho^D_{AS}(x,x)+\rho^D_{SA}(x,x)+\rho^D_{AA}(x,x)$ , we arrive at
\begin{eqnarray}
\rho^D_S(z) = (1-2\epsilon )\rho^D_S(2z) + g(z)
\end{eqnarray} 
where $g(z) =  (1-2\epsilon)\rho^D_A(2z)$. Its solution can
be written down as
\begin{eqnarray}
\rho^D_S(z) = \sum_{k=0}^{\infty} (1-2\epsilon)^k g(2^kz).
\end{eqnarray}
This is a Weierstrass function and if
$g(z)\in C^{\beta}$ where $\beta > -\ln(1-2\epsilon)/\ln 2$
then the calculation in~\cite{falconer1990} for $g(z) = \sin(z)$
can be carried over and it can be shown that the pointwise 
H\"older exponent of this function is $-\ln(1-2\epsilon)/\ln 2$
everywhere implying that ${\rm dim_Bgraph}\rho^D_S =d$.
And if $g(z)$ is not smooth enough then it can only
increase the box dimension  of $\rho^D(x)$, hence the
result. \qed

\section{Concluding Remarks}\label{sec:4}
We have studied the stationary density of two coupled tent maps
as a function of the coupling parameter.
We find that even though the density of the individual tent
map is  smooth it becomes very irregular as soon as a small
coupling is introduced in the sense that the pointwise
H\"older exponent is small everywhere. 
And the density smoothes as the coupling
is increased. It becomes smooth that is the 
H\"older exponent becomes one at the value of $\epsilon$
where the synchronization transition takes place.
We have thus elucidated a new aspect of synchronization
in coupled dynamical systems, beyond the standard aspects of  
linear or global stability of synchronized solutions.

It is a curious fact that the H\"older exponent becomes
one exactly at the critical value of the coupling 
parameter, i.e.,  $\epsilon_c = 1/4$. 
It would be important to understand 
if there is any underlying principle behind this
observation, that is, one valid also for other maps
with varying coupling matrices.

It is interesting to note that fractal probability 
densities have arisen in a completely different scenario, namely 
the random walk problem with shrinking step 
lengths~\cite{Krapivsky2004}.

One should also characterize the stationary density away
from the synchronization manifold. It is expected to have a 
more complex multifractal character~\cite{Jaffard1997}.
The effect of different network topologies on the
stationary density is another interesting topic. 

One of us (KMK) would like to thank the Alexander-von-Humboldt-Stiftung for 
financial
support.

%
\bibliographystyle{unsrt}
\bibliography{kolwankar}

\begin{thebibliography}{10}

\bibitem{PRK}
A.~Pikovsky, M.~Rosenblum, and J.~Kurths.
\newblock {\em Synchronization - A Universal Concept in Nonlinear Science}.
\newblock Cambridge University Press, 2001.

\bibitem{Pecora1997}
L.~M. Pecora, T.~L. Carroll, G.~A. Johnson, D.~J. Mar, and J.~F. Heagy.
\newblock Fundamentals of synchronization in chaotic systems, concepts, and
  applications.
\newblock {\em Chaos}, 7:520, 1997.

\bibitem{VanWiggeren1998}
G.~D. VanWiggeren and R.~Roy.
\newblock Communication with chaotic lasers.
\newblock {\em Science}, 279:1198, 1998.

\bibitem{Hansel1992}
D.~Hansel and H.~Sompolinsky.
\newblock Synchronization and computation in a chaotic neural network.
\newblock {\em Phys.\ Rev.\ Lett.}, 68:718, 1992.

\bibitem{Buzsaki2004}
G.~Buzs\'aki and A.~Draguhn.
\newblock Neuronal oscillations in cortical networks.
\newblock {\em Science}, 304:1926, 2004.

\bibitem{Friedrich2004}
R.~W. Friedrich, C.~J. Habermann, and G.~Laurent.
\newblock Multiplexing using synchrony in the zebrafish olfactory bulb.
\newblock {\em Nature Neuroscience}, 7:862, 2004.

\bibitem{Jost2004}
J.~Jost and K.~M. Kolwankar.
\newblock Global analysis of synchronization in coupled maps, 2004.

\bibitem{jost2002}
J.~Jost and M.~P. Joy.
\newblock Spectral properties and synchronization in coupled map lattices.
\newblock {\em Phys. Rev. E}, 65(1):016201, 2002.

\bibitem{LM}
A.~Lasota and M.~C. Mackey.
\newblock {\em Chaos, Fractals and Noise}.
\newblock Springer, 1994.

\bibitem{falconer1990}
K.~Falconer.
\newblock {\em Fractal Geometry - Mathematical Foundations and Applications}.
\newblock John Wiley, 1990.

\bibitem{Krapivsky2004}
P.~L. Krapivsky and S.~Redner.
\newblock Random walk with shrinking steps.
\newblock {\em Am. J. Phys.}, 72:591, 2004.

\bibitem{Jaffard1997}
S.~Jaffard.
\newblock Multifractal formalism for functions part ii: Self-similar functions.
\newblock {\em SIAM J. Math. Anal.}, 28:971, 1997.

\end{thebibliography}

\printindex
\end{document}